\documentstyle[12pt,twoside,fleqn,esppre,epsfig]{article}
\newcommand{\nslash}{{\rm n} \hskip -0.5em /}
\newcommand{\dslash}{\partial \hskip -0.5em /}
\newcommand{\aslash}{a \hskip -0.5em /}
\newcommand{\be}{\begin{eqnarray}}
\newcommand{\ee}{\end{eqnarray}}

\newcommand{\AmS}{{\protect\the\textfont2
  A\kern-.1667em\lower.5ex\hbox{M}\kern-.125emS}}

\title{Nucleon Structure Functions in a Chiral
       Soliton Model\thanks{Talk presented at the KEK--Tanashi
       Int. Symposium on the {\it Physics of Hadrons and Nuclei},
       Tokyo, Dec. 1998, to appear in the proceedings.
       This work is supported in parts by funds
       provided by the U.S. Department of Energy (D.O.E.)
       under cooperative research agreement
       \#DF--FC02--94ER40818 and by the
       Deutsche Forschungsgemeinschaft (DFG) under 
       contract We 1254/3-1.}}%

\author{Herbert Weigel\address{Center for Theoretical Physics\\
       Laboratory of Nuclear Science and Department of Physics\\
       Massachusetts Institute of Technology,
       Cambridge, Ma 02139, USA}%
        \thanks{Heisenberg Fellow}%
}

\begin{document}
\maketitle

\begin{abstract}
The computation of nucleon structure functions within the 
Nambu--Jona--Lasinio chiral soliton model is outlined. After
some technical remarks on the issue of regularization numerical
results for the both unpolarized and polarized structure functions
are presented. The generalization to flavor SU(3) is sketched.
\end{abstract}

\section{THE CHIRAL SOLITON MODEL}
The bosonized form of the Nambu--Jona--Lasinio (NJL) action
\cite{Na61}
\be
{\cal A} [S,P]=-iN_C{\rm Tr}_{\textstyle\Lambda}{\rm log}\, 
\left[i\dslash-\left(S+i\gamma_5P\right)\right]
-\frac{1}{4G}\int d^4x\, {\rm tr}\,
\left[S^2+P^2+2{\hat m}_0(S+iP)\right]\, ,
\label{bosact}
\ee
represents the starting point for considering nucleon structure 
functions in a chiral soliton model. The action ${\cal A}$
is a functional of respectively  scalar and pseudoscalar fields 
$S$ and $P$ which are matrices in flavor space. 
The VEV, $\langle S\rangle=m$ 
is obtained from the gap--equation and 
measures the dynamical breaking of chiral symmetry. For apparent 
reasons $m$ is called the constituent quark mass (matrix). The 
regularization of the quadratically divergent quark loop is indicated 
by the cut--off $\Lambda$. Its value as well as the coupling 
constant $G$ and the current quark mass ${\hat m}_0$ are adjusted to 
the phenomenological meson parameters $m_\pi$ and $f_\pi$,
leaving only a single free parameter, commonly chosen to 
be $m$. The static soliton is constructed from the hedgehog {\it ansatz} 
on the chiral circle 
\be
S+i\gamma_5 P = m\, {\rm exp}\left(
i{\vec \tau}\cdot {\hat r} \gamma_5 \Theta(r)\right)
=: m\, U_5\, .
\label{hedgehog}
\ee
This defines a Dirac Hamiltonian 
$h={\vec\alpha}\cdot{\vec p} + m\, \beta\, U_5$ with eigenvalues 
$\epsilon_\alpha$. The latter parameterize the 
regularized energy functional extracted from the action (\ref{bosact})
\be
E[\Theta]=N_C\left(\epsilon_{\rm val}- \frac{1}{2}
{\sum_\alpha}_{\textstyle\Lambda} 
\left|\epsilon_\alpha\right|\right)\, .
\label{engfct}
\ee
Here the subscript ``${\rm val}$'' refers to the distinct valence quark 
level which is strongly bound in the background of the hedgehog. 
The soliton is finally constructed by extremizing the functional 
(\ref{engfct}). Details of this approach and the computation of 
static nucleon properties are extensively discussed in the 
literature \cite{Al96}.

\section{REMARKS ON REGULARIZATION}
DIS off the nucleon is parameterized by 
the hadronic tensor $W^{\mu\nu}(q)$ with $q$ being the momentum 
transmitted to the nucleon. $W^{\mu\nu}(q)$ is obtained from the 
nucleon matrix element of the commutator 
$[J^\mu(\xi), J^\nu(0)]$. In the NJL model the current is conveniently 
given as $J^\mu={\bar q}{\cal Q}\gamma^\mu q$, with
${\cal Q}$ being the quark charge matrix. In the context of functional 
bosonization it is more appropriate to start from the forward virtual 
Compton amplitude,
\be
T^{\mu\nu}(q)=\int d^4x\, {\rm e}^{iq\cdot\xi}\,
\langle N|T\left(J^\mu(\xi) J^\nu(0)\right)|N\rangle\, ,
\label{comp1}
\ee
since the time--ordered 
product is unambiguously extracted from the regularized action 
\be
T\left(J^\mu(\xi) J^\nu(0)\right)=
\frac{\delta^2}{\delta a_\mu(\xi)\, \delta a_\nu(0)}\,
{\rm Tr}_\Lambda {\rm log}\,
\left[i\dslash-\left(S+i\gamma_5P\right)+{\cal Q}\,
\aslash\right]\Big|_{a_\mu=0}\, \, .
\label{tprod}
\ee
In this way the regularization of the structure functions is 
consistently implemented at the level of the defining action.
The hadronic tensor is then obtained from the cut,
\be
W^{\mu\nu}(q)=\frac{1}{2\pi} \Im\, (T^{\mu\nu}(q))\, .
\label{abspart}
\ee
In order to extract the leading twist piece of the structure 
functions, $W^{\mu\nu}(q)$ in studied in the Bjorken limit: 
$q^2\to-\infty$ with $x=-q^2/P\cdot q$ fixed. Here $P$ denotes the 
nucleon momentum. In this limit the leading order contribution in 
$1/N_C$ to $W^{\mu\nu}(q)$ becomes \cite{Wexx}
\be
W^{\mu\nu}(q)&=&i\frac{N_C}{4}
\int \frac{d\omega}{2\pi}\sum_\alpha \int d^3 \xi
\int \frac{d\lambda}{2\pi}\, {\rm e}^{iMx\lambda}
\nonumber \\ && \hspace{-2.3cm}
\times\Bigg\{\Big[\Psi^\dagger_\alpha({\vec\xi}\,){\cal Q}^2
\gamma^\mu\nslash\gamma^\nu\beta
\Psi_\alpha({\vec\xi}+\lambda{\hat e}_3)
{\rm e}^{-i\lambda\omega}
\hspace{-0.1cm}-\hspace{-0.1cm}
\Psi^\dagger_\alpha({\vec\xi}\,){\cal Q}^2\beta
\gamma^\nu\nslash\gamma^\mu\Psi_\alpha({\vec\xi}+\lambda{\hat e}_3)
{\rm e}^{i\lambda\omega}\Big]f_\alpha^{(-)}(\omega)_{\rm pole}
\nonumber \\ && \hspace{-2.2cm}
+\Big[\Psi^\dagger_\alpha({\vec\xi}\,){\cal Q}^2\beta
\gamma^\mu\nslash\gamma^\nu\Psi_\alpha({\vec\xi}+\lambda{\hat e}_3)
{\rm e}^{-i\lambda\omega}
\hspace{-0.1cm}-\hspace{-0.1cm}
\Psi^\dagger_\alpha({\vec\xi}\,){\cal Q}^2
\gamma^\nu\nslash\gamma^\mu\beta
\Psi_\alpha({\vec\xi}+\lambda{\hat e}_3)
{\rm e}^{i\lambda\omega}\Big]f_\alpha^{(+)}(\omega)_{\rm pole}
\Bigg\}\, ,
\label{hadten}
\ee
with the Pauli--Villars regularized spectral functions,
\be
f_\alpha^{(\pm)}(\omega)=
\sum_i c_i \frac{\omega\pm\epsilon_\alpha}
{-\omega^2+\epsilon_\alpha^2+\Lambda_i^2-i\eta}
\pm\frac{\omega\pm\epsilon_\alpha}
{-\omega^2+\epsilon_\alpha^2-i\eta}\, ,
\label{specfct}
\ee
and ${\rm n}^\mu=(1,0,0,1)$ being a light--cone vector.
As there are poles for both positive and negative $\omega$, the 
meaning of forward and backward moving quarks becomes ambiguous 
and a description of $W_{\mu\nu}$ in terms of quark distributions 
seems impossible.

\section{NUMERICAL RESULTS}
Unfortunately no numerical results are currently available for the 
structure functions as projected from the consistently regularized 
hadronic tensor (\ref{hadten}). Therefore the presentation is 
limited to the valence quark approximation taking into account only 
the contribution of the distinct valence level. Two observations 
make this a reliable approximation: (1) This level does not undergo 
regularization and hence the subtleties mentioned above are avoided.
(2) Although the polarized vacuum is mandatory to provide a 
soliton solution, its contribution to static properties is 
small or negligible \cite{Al96}. Sum rules relate them to 
structure functions, thus it is suggestive that the structure 
functions are also saturated by the valence quark contribution. 
For more details the reader is referred to the research papers 
\cite{We96,We97,Sch98} and similar studies by other groups \cite{Di96,Wa98}.

\subsection{Unpolarized Structure Functions}
The unpolarized structure functions are obtained from the
symmetric combination $W_{\mu\nu}+W_{\nu\mu}$. 
The result for the structure function which enters
the Gottfried sum rule of $e-N$ scattering is shown 
in fig. 1. The boost to the infinite momentum frame  
\cite{Ga98} mitigates the effects of omitting the dynamical 
response of the soliton to the infinite momentum transfer and
provides proper support for the structure functions. 
A DGLAP evolution \cite{DGLAP} determines the low energy 
scale $Q_0^2\approx0.4{\rm GeV^2}$ at which the model supposedly 
approximates QCD. The model reproduces the gross features
of the experimental data although an even better 
agreement can be gained by further reducing $Q_0^2$ which, however,
would make the DGLAP program unrealistic.
Depending on the model parameter $m$, the Gottfried sum
rule, $S_G=\int dx (F_2^{\rm ep}-F_2^{\rm en})/x$ is found to be
0.26--0.29 which exhibits the desired deviation from the
historic value (1/3) demanded empirically,
$0.235\pm0.026$ \cite{NMC94}.

\bigskip
\parbox[l]{5.5cm}
{{\sf Fig.~1: The unpolarized structure function
entering the Gott\-fried sum rule. RF: rest frame, IMF: boosted
to the infinite momentum frame, LO: leading order QCD evolution 
to $Q^2=4{\rm GeV^2}$. Data are from the NMC \cite{NMC94}.}}
\parbox[r]{10.0cm}
{\centerline{\hspace{1cm}
\epsfig{figure=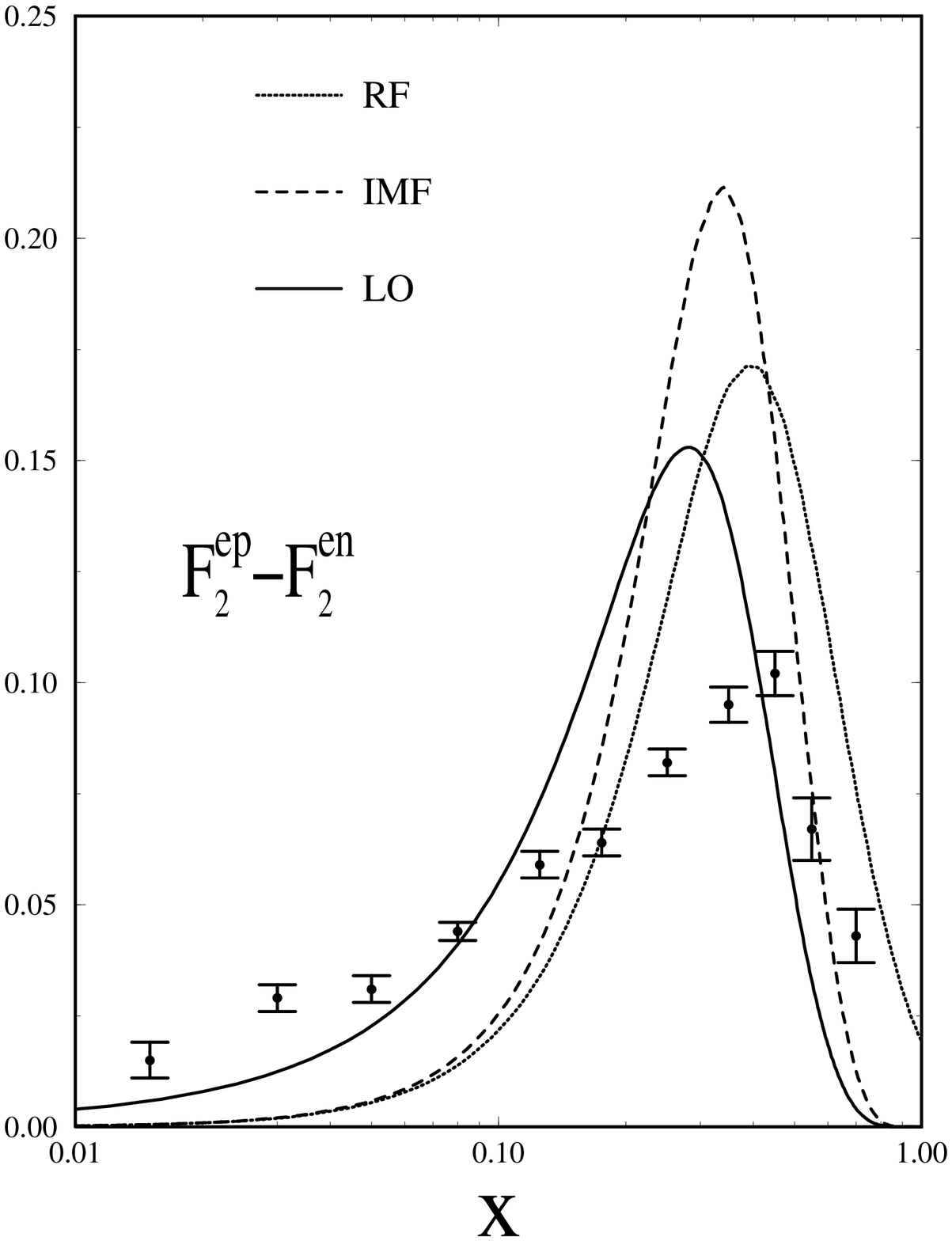,height=4.0cm,width=9.0cm}}
}
\bigskip

\subsection{Spin Structure Functions}
The spin -- polarized structure functions $g_1$ and $g_2$ are obtained 
from the anti--symmetric combination $W_{\mu\nu}-W_{\nu\mu}$. 
While the former is related to the proton spin puzzle, 
the latter sheds some light on higher twist effects \cite{Ja95}. 
Typical results are shown in fig. 2 and are compared to data from 
SLAC \cite{Abe95}. Apparently a reasonable agreement is obtained.
\pagebreak
\begin{center}
{\sf Fig. 2: Model predictions for the polarized 
nucleon structure functions $g_1$ (left panel) \\
and $g_2$ (right panel). Data are from SLAC \cite{Abe95}. }
\bigskip
\bigskip

\epsfig{figure=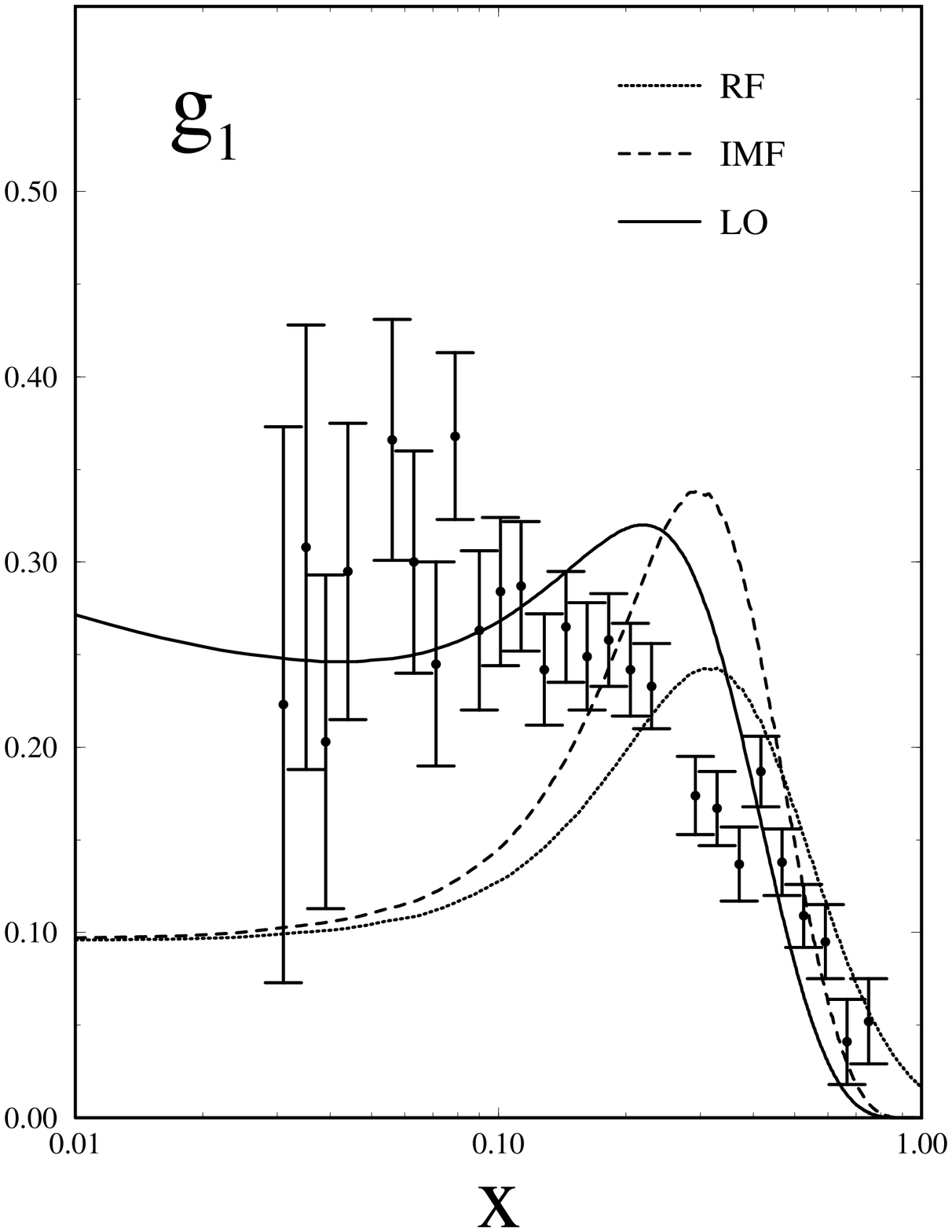,height=4.3cm,width=7.0cm}
\hspace{1cm}
\epsfig{figure=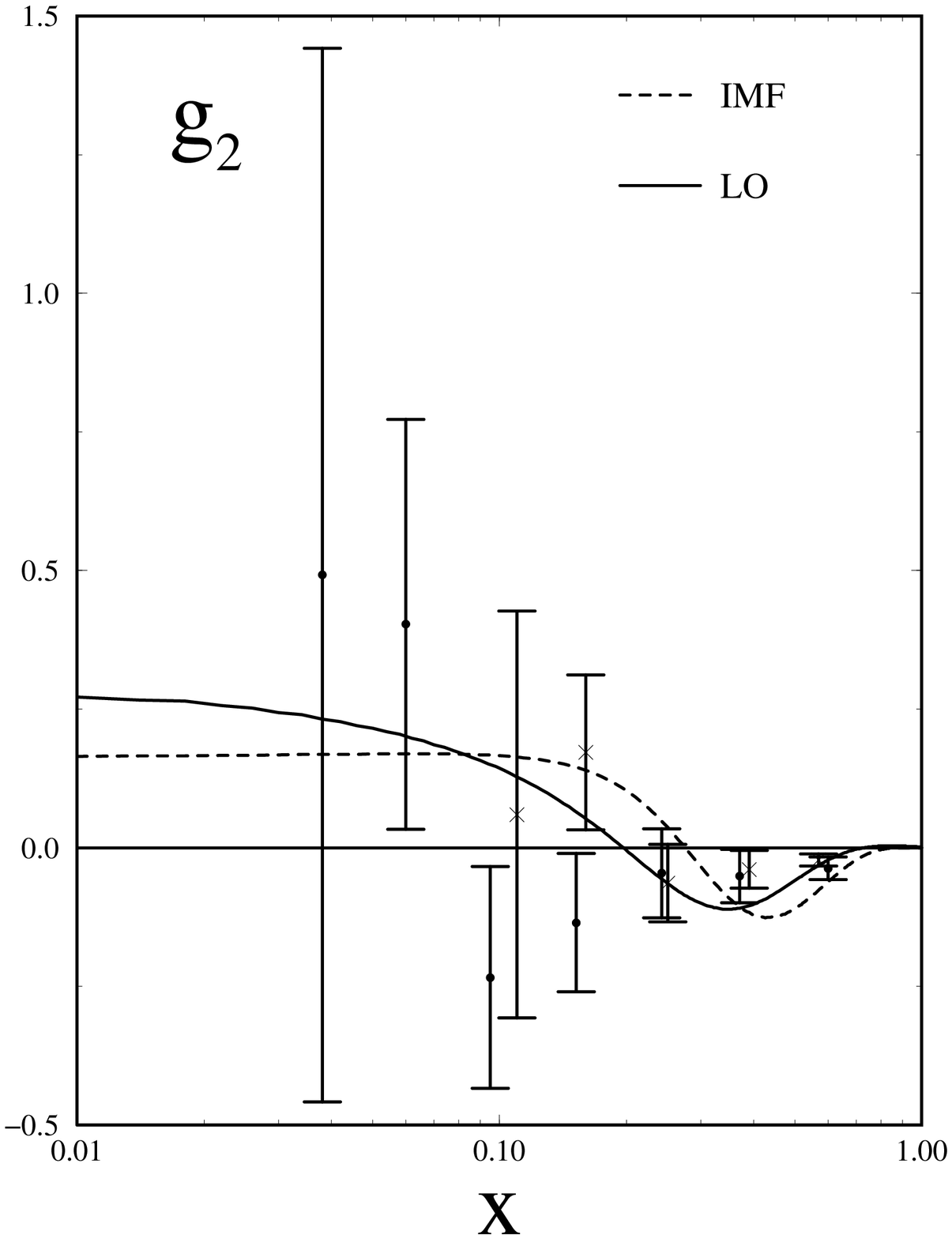,height=4.3cm,width=7.0cm}
\end{center}
\bigskip

\subsection{Generalization to Flavor SU(3)}
The model can straightforwardly be generalized to also include strange 
quarks through the collective coordinate quantization \cite{Sch98}. 
In a flavor symmetric formulation one would expect sizable strange
quark contributions to all nucleon properties. Fortunately flavor 
symmetry breaking effects can be included, thereby considerably 
reducing the strange quark contribution to $g_1$, {\it cf.} fig. 3. 
This is also reflected by the small difference between the two
and three flavor model calculation for $g_1$ of the proton.
\begin{center}
{\sf Fig.~3: Left panel: Strangeness contribution to $g_1$.
Right panel: Comparison of the two and three flavor model
calculation of $g_1$ of the proton. Data are from ref \cite{Abe95}.}
\bigskip
\bigskip

\epsfig{figure=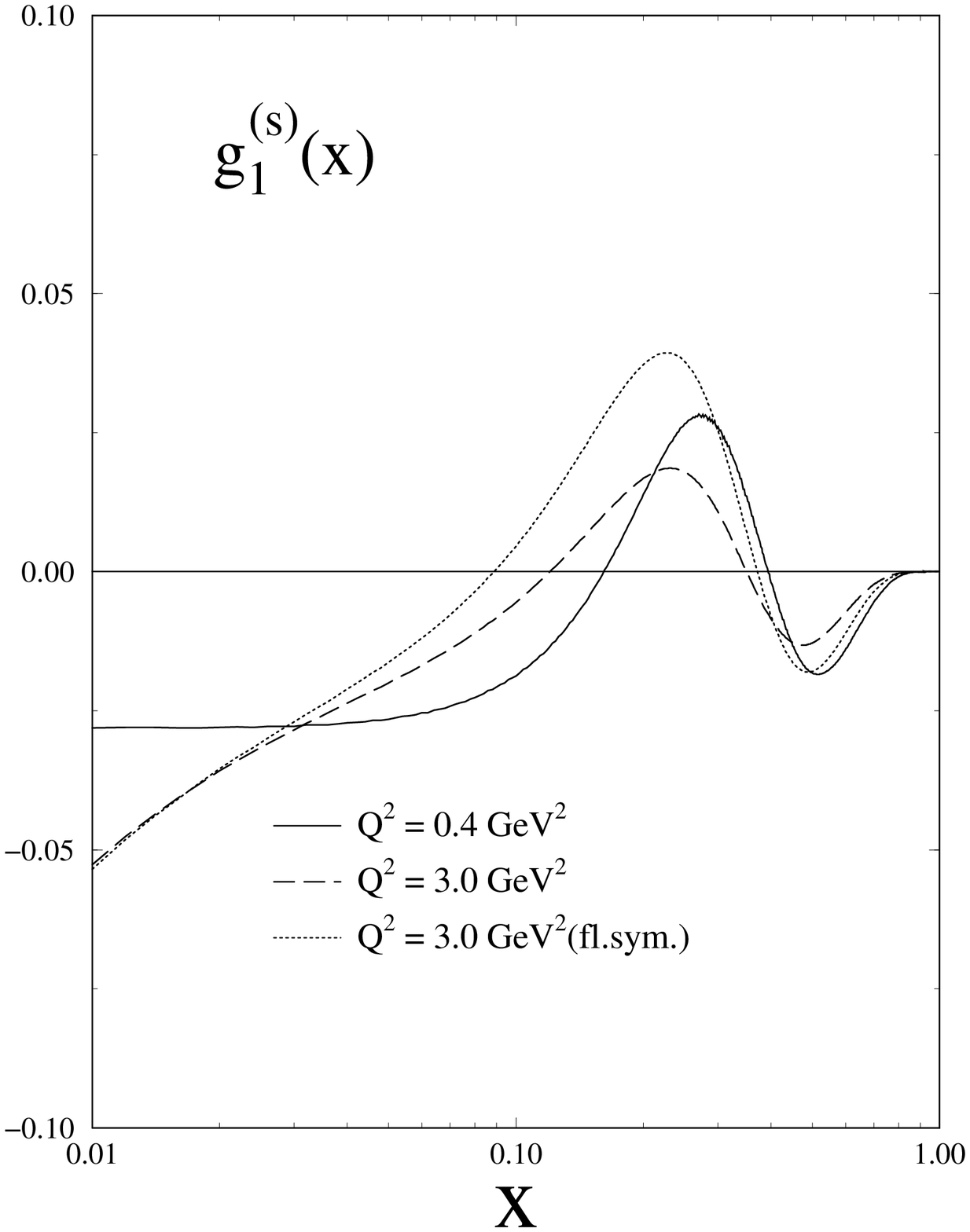,height=4.8cm,width=7.0cm}
\hspace{1cm}
\epsfig{figure=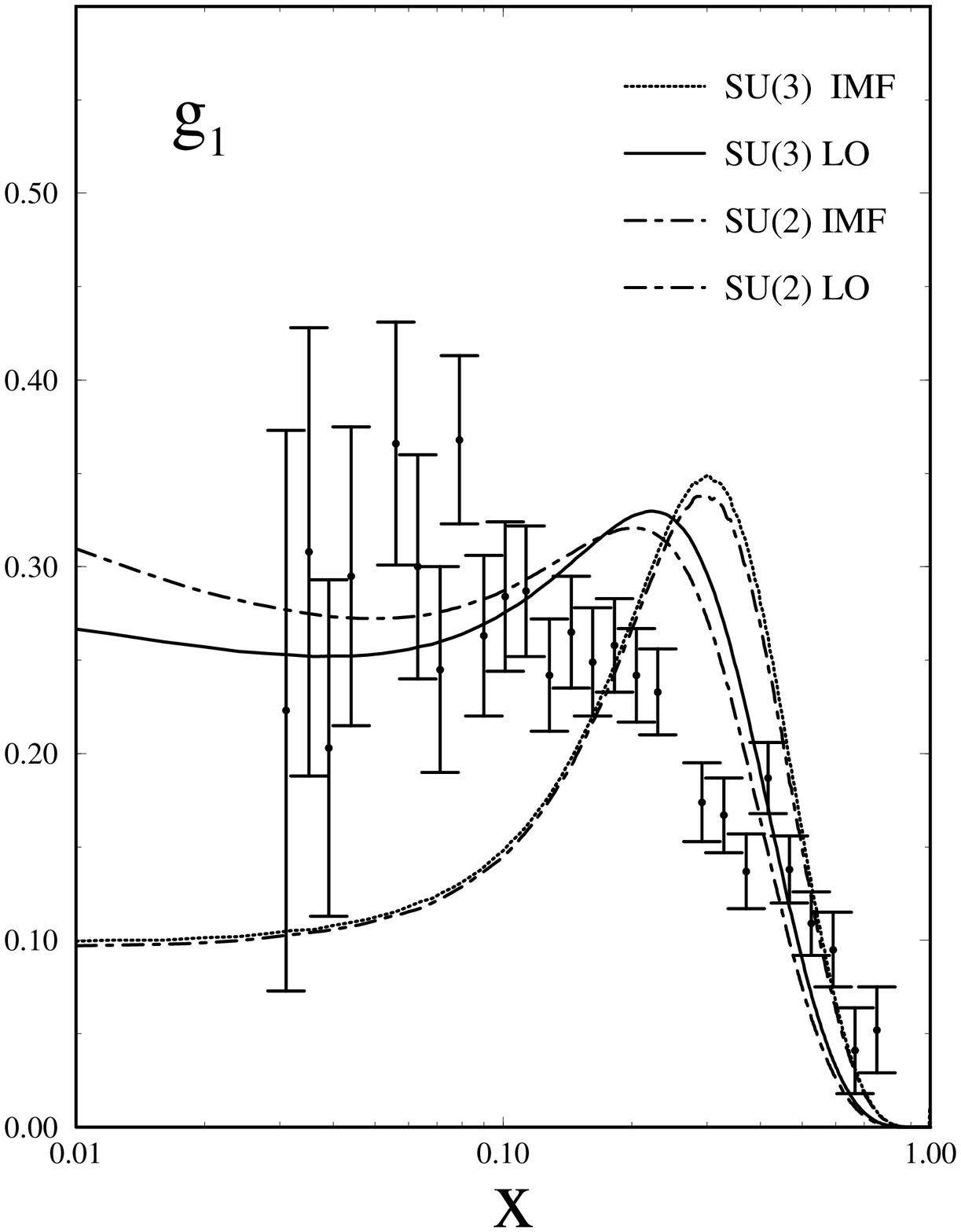,height=4.8cm,width=7.0cm}
\end{center}
\bigskip

\section{CONCLUSIONS}
Nucleon structure functions can be computed from a chiral 
soliton model which des\-cribes baryons as lumps
of mesons. All relevant information is contained
in the hadronic tensor which can be computed from the 
symmetry currents of the model. Thus  the identification of 
degrees of freedom with those of QCD, which seems impossible 
after regularization, is unnecessary. Reasonable agreement
with experimental data is obtained in the valence quark
approximation which leaves aside the technical subtleties of 
regularization.

\section*{ACKNOWLEDGMENTS}
It is my pleasure to thank the organizers for providing
a very pleasant and memorable symposium. The material
presented stems from collaborations with L. Gamberg,
H. Reinhardt, O. Schr\"oder and E. Ruiz Arriola whose 
contributions are gratefully acknowledged.


\begin{thebibliography}{9}
\small
\bibitem{Na61}
Y. Nambu and G. Jona--Lasinio,
\newblock Phys. Rev. {\bf 122} (1961) 345; {\bf 124} (1961) 246,\\
D. Ebert and H. Reinhardt,
\newblock Nucl. Phys. {\bf B271} (1986) 188.
\bibitem{Al96}
R. Alkofer, H. Reinhardt and H. Weigel,
\newblock Phys. Rep. {\bf 265} (1996) 139,\\
C. V. Christov {\it et al.}, Prog. Part. Nucl. Phys. {\bf 37} (1996) 91.
\bibitem{Wexx}
H. Weigel, E. Ruiz Arriola and L. Gamberg,
in preparation.
\bibitem{We96}
H. Weigel, L. Gamberg and H. Reinhardt,
Mod. Phys. Lett. {\bf A11} (1996) 3021,\
Phys. Lett. {\bf B399} (1997) 28.
\bibitem{We97}
H. Weigel, L. Gamberg and H. Reinhardt,
\newblock Phys. Rev. {\bf D55} (1997) 6910,\\
L. Gamberg, H. Reinhardt and H. Weigel,
Phys. Rev. {\bf D58} (1998) 054014.
\bibitem{Sch98}
O. Schr\"oder, H. Reinhardt and H. Weigel,
Phys. Lett. {\bf B439} (1998) 398, \
hep--ph/9902322.
\bibitem{Di96}
D. I. Diakonov {\it et al.},
Nucl. Phys. {\bf B480} (1996) 341,\
Phys. Rev. {\bf D56} (1997) 4069.
\bibitem{Wa98}
M. Wakamatsu and T. Kubota,
Phys. Rev. {\bf D57} (1998) 5755,\
hep--ph/9809443.
\bibitem{Ga98}
L. Gamberg, H. Reinhardt and H. Weigel,
Int. J. Mod. Phys. {\bf A13} (1998) 5519.
\bibitem{DGLAP}
V. N. Gribov and L. N. Lipatov,
Sov. J. Nucl. Phys. {\bf 15} (1972) 438,\\
Y. L. Dokshitzer, Sov. Phys. JETP {\bf 46} (1977) 461,  \\
G. Altarelli and G. Parisi,
\newblock Nucl. Phys. {\bf B126} (1977) 298.
\bibitem{NMC94}
M. Arneodo {\it et al.} (NMC),
Phys. Rev. {\bf D50} (1994) R1.
\bibitem{Ja95}
R. L. Jaffe, ``Spin, Twist and Hadron Structure...''
Erice Lectures 1995, hep--ph/9602236.
\bibitem{Abe95}
K.\ Abe {\it et al.},
Phys. Rev. Lett. {\bf 74} (1995) 346.
\end{thebibliography}
\end{document}